\documentstyle[manuscript,aps,prl]{revtex}

\input{epsf}

\begin{document}

\title{Entangled-photon Fourier optics}

\author{Ayman F. Abouraddy, Bahaa E. A. Saleh, Alexander V. Sergienko, and Malvin C. Teich}
\address{Quantum Imaging Laboratory, Departments of Electrical $\&$ Computer Engineering and Physics, Boston University, Boston, MA $02215-2421$}

\maketitle
\begin{abstract}
Entangled photons, generated by spontaneous parametric
down-conversion from a second-order nonlinear crystal, present a
rich potential for imaging and image-processing applications.
Since this source is an example of a three-wave mixing process,
there is more flexibility in the choices of illumination and
detection wavelengths and in the placement of object(s) to be
imaged. Moreover, this source is entangled, a fact that allows for
imaging configurations and capabilities that cannot be achieved
using classical sources of light. In this paper we examine a
number of imaging and image-processing configurations that can be
realized using this source. The formalism that we utilize
facilitates the determination of the dependence of imaging
resolution on the physical parameters of the optical arrangement.
\end{abstract}


\section{INTRODUCTION}

The process of optical parametric three-wave mixing in a
second-order nonlinear medium
\cite{JA,DKleinman,RK,SA,NB,FZ,AY,PB,BS1,DM} involves the coherent
interaction between three optical fields with, generally,
different wavelengths: pump, signal, and idler. Because of the
phase-matching requirements \cite{FZ,AY,DM}, the wave-vectors are
related and the spatial distributions of the fields are therefore
highly coupled. This process may therefore be utilized in {\it
distributed multi-wavelength imaging} or image processing, where
objects are placed in the path of one or more of these fields of
different wavelengths and the intensities or cross-correlations
are measured. Two examples of phenomena based on three-wave
mixing, optical parametric oscillation (OPO) and optical
parametric amplification (OPA), have been studied extensively and
many interesting phenomena of spatial correlation \cite{AG},
pattern formation \cite{AB1,LL}, and reduced-noise image
amplification \cite{SC,IS} have been reported.

A third example is the process of spontaneous parametric
down-conversion (SPDC) \cite{DK1}, a phenomenon that exhibits
quantum entanglement \cite{ES}. The signal and idler waves,
created when the nonlinear medium is illuminated by an intense
laser beam (pump), are produced in the form of photon pairs in an
entangled quantum state (biphotons). Spatial and spectral
entanglement are a consequence of the multiple possibilities for
satisfying conservation of energy and momentum for each photon
pair. Interest in the SPDC process has spurred many studies of its
spatial and spatio-temporal photon correlation properties
\cite{DK2,AM,DK3,PR1,AJ1,DS,AJ2,AB2,CM,BM,AA1}, and some imaging
applications based on the measurement of photon-pair coincidence
have been proposed \cite{AB3,BS2} and tested \cite{TP1,TP2}.

In this paper we develop a general Fourier-optics theory of image
formation based on the SPDC process. In Section 2 we explore new
configurations for multi-wavelength distributed imaging and
image-processing applications. We follow an approach introduced in
a previous paper \cite{BS3}, in which we established a duality
between partial quantum entanglement and classical partial
coherence theory. We use the formalism developed in Ref.
[\cite{BS3}] and apply it to the image-formation process in
distributed multi-wavelength imaging configurations made possible
by the nature of this SPDC source. In Section 3 we study the
imaging resolution of these configurations and the effect on it of
the various physical parameters of the system. In the Appendix we
provide a brief review of classical imaging theory in the
framework of the optical bilinear transformation.

\section{CONFIGURATIONS FOR SPDC BIPHOTON IMAGING AND IMAGE
PROCESSING}

The principal function of an optical imaging system is to transfer
the spatial distribution of some physical property of an object
(transmittance, reflectance, or absorbance), via an optical wave,
to a remote location where it is measured with a photodetector
\cite{BS1,MB}. An image-processing system transforms one image
into another with enhanced features, or obtains a new image from
more than one image, such as the correlation of two images
\cite{JG,GR,FY}. We examine here various configurations for
imaging and image processing based on biphoton beams generated by
SPDC. As mentioned above, the existence of three optical beams
(pump, signal, and idler) allows us to construct imaging
configurations that are not achievable with other single-beam
optical sources. We may place an object that is to be imaged in
either of the three beams. We may alternatively place more than
one object in these beams and obtain the correlation of their
images.

All these configurations include two features. The first is that
they are examples of distributed imaging \cite{AB3,BS2,AA2}. In
analogy to distributed computing, where the computation resources
are distributed over a network, distributed imaging allows us to
reallocate the imaging components from the particular path
connecting the source to the object to be imaged. The second is
that they allow the possibility of multi-wavelength imaging: the
object may be illuminated with one wavelength whereas detection
takes place at another.

The use of other nonclassical sources of light in imaging has been
shown to lead to noise reduction \cite{SC,IS,MK1,MK2}. In this
paper we direct our attention to various {\it imaging
configurations}. The quantum nature of the SPDC source offers the
additional advantage of noise reduction, but this is immaterial to
the task at hand.

In the process of SPDC, an intense laser beam (pump) illuminates a
nonlinear crystal (NLC) with quadratic nonlinear susceptibility
\cite{AY,DK1}. Some of the pump photons disintegrate into pairs of
photons (known traditionally as signal and idler), which conserve
the energy and momentum of the parent pump photon. Consider the
situation depicted in Fig. \ref{overall layout}. The pump beam
illuminates the NLC and the signal and idler beams are measured by
the single-photon detectors D$_1$ and D$_2$, respectively. We
assume throughout a planar source and a one-dimensional geometry
in the transverse plane for the sake of simplicity but without
loss of generality. Optical systems, containing objects to be
imaged and any optical components, may be placed in any of the
three available beams.

The signal and idler photons can be emitted from the NLC in a
variety of configurations. The two photons may be emitted in two
different and distinct directions, in which case each photon will
pass through a different (and possibly remote) optical system;
this configuration is denoted non-collinear. The two photons may
be emitted in the same spatial wave packet, the collinear case,
but have some distinguishing characteristic, such as frequency or
polarization, whereupon they effectively pass through different
optical systems if the components are dispersive or polarization
dependent. The two photons are detected in the same output plane
in this case. Finally, the two photons may be emitted in the same
spatial wave packet and have no distinguishing characteristic, the
collinear degenerate case, and thus they pass through the same
optical system and are detected in the same output plane.

The coincidence rate of photon pairs at the two detectors, D$_1$
and D$_2$ located at positions $x_{1}$ and $x_{2}$, respectively,
is proportional to the fourth-order correlation function of the
fields, $G^{(2)}(x_{1},x_{2})$ \cite{BS3,RG}, the biphoton rate.
The signal and idler beams traverse optical systems described by
their impulse response functions $h_{s}(x_{1},x)$ and
$h_{i}(x_{2},x)$, respectively. It has been shown that the
biphoton rate is given by \cite{BS3}
\begin{equation}\label{G2psi}
G^{(2)}(x_{1},x_{2})=|\psi(x_{1},x_{2})|^{2},
\end{equation}
where the biphoton amplitude $\psi(x_{1},x_{2})$ is
\begin{equation}\label{basicEq}
\psi(x_{1},x_{2})=\int dxE_{p}(x)h_{s}(x_{1},x)h_{i}(x_{2},x);
\end{equation}
here $E_{p}(x)$ is the spatial distribution of the pump field at
the entrance to the NLC. The result in Eq. (\ref{basicEq}) was
derived assuming a thin NLC and the presence of narrowband
spectral filters in the optical system. These two assumptions
simplify the analysis considerably without overshadowing the
physics of the imaging processes discussed. They will be relaxed
in Section 3.

An interpretation of Eq. (\ref{basicEq}) that is useful in
understanding the behavior of such a system was advocated by
Klyshko \cite{DK2,DK3,DS,AB3} under the name "advanced wave
interpretation". In this picture, the biphoton amplitude in Eq.
(\ref{basicEq}) can be viewed as the impulse response function of
an optical system represented by the cascade of three systems:
propagation from D$_{1}$ at $x_{1}$ back through a system with
impulse response function $h_{s}^{r}(x,x_{1})=h_{s}(x_{1},x)$,
modulation by the pump field $E_{p}$, and subsequent transmission
through a system with impulse response function $h_{i}(x_{2},x)$.
An intuitive advantage can be gained by unfolding the system in
this way as will become clear shortly.

Two special correlation functions deriving from
$G^{(2)}(x_{1},x_{2})$ in Eq. (\ref{G2psi}) are of interest: the
marginal coincidence rate $I^{(2)}(x_{2})$, and the conditional
coincidence rate $I_{0}^{(2)}(x_{2})$, defined by
\begin{equation}\label{I2x}
I^{(2)}(x_{2})=\int dx_{1}G^{(2)}(x_{1},x_{2}),
\end{equation}
\begin{equation}\label{I20x}
I_{0}^{(2)}(x_{2})=G^{(2)}(0,x_{2}).
\end{equation}
The marginal coincidence rate $I^{(2)}(x_{2})$ is proportional to
the probability of detecting a photon at $x_{2}$ by D$_{2}$ when
detector D$_{1}$ detects a photon at any location
$-\infty<x_{1}<\infty$. The conditional coincidence rate
$I_{0}^{(2)}(x_{2})$ is proportional to the probability of
detecting a photon at $x_{2}$ by D$_{2}$ when D$_{1}$ detects a
photon at $x_{1}=0$.

We proceed to examine the five configurations that are possible
with this optical source and explore their imaging and
image-processing potential.

\subsection{Object in the signal (or idler) beam}

The generalized biphoton optical system described by Eq.
(\ref{basicEq}) permits the object to be placed in {\it either}
the signal {\it or} the idler beams such that its transmittance
(or reflectance) modifies either of the impulse response functions
$h_{s}$ or $h_{i}$. Without loss of generality, we assume that the
object is placed in the signal beam. However, the choice of either
beam might be dictated by wavelength considerations.

Consider the situation depicted in Fig. \ref{object in the signal
beam}. The reverse signal system
$h_{s}^{r}(x,x_{1})=h_{s}(x_{1},x)$  is regarded as a cascade of
two linear systems of impulse response functions $h_{1}$ and
$h_{2}$ with the object $t$ sandwiched in between. The biphoton
amplitude can thus be written as
\begin{equation}\label{basicEqObjInSignal}
\psi(x_{1},x_{2})=\int dx't(x')h_{1}(x_{1},x')h_{3}(x_{2},x'),
\end{equation}
where $h_{3}$ is the impulse response function of a system
composed of a cascade of the reverse of system $h_{2}$, an
aperture $E_{p}$, and the system $h_{i}$, and is given by
\begin{equation}\label{h3}
h_{3}(x_{2},x')=\int dxE_{p}(x)h_{i}(x_{2},x)h_{2}(x',x).
\end{equation}
Equation (\ref{basicEqObjInSignal}) states that the overall system
is composed of an illumination system $h_{1}$ illuminating the
object $t$, followed by an imaging system $h_{3}$, which is
dependent on $h_{2}$, $h_{i}$, and $E_{p}$, in accordance with Eq.
(\ref{h3}). The uniformity of the illumination system $h_{1}$ and
the resolution of the system $h_{3}$ determine the quality of the
overall imaging system.

In this configuration, then, the conditional coincidence rate,
obtained using Eqs. (\ref{G2psi}), (\ref{I20x}), and
(\ref{basicEqObjInSignal}), is
\begin{equation}\label{I20ObjInSig}
I_{0}^{(2)}(x_{2})=|\int dx't(x')h_{1}(0,x')h_{3}(x_{2},x')|^{2}.
\end{equation}
This system is mathematically equivalent to a coherent optical
system where the object is modulated by the illumination
distribution $h_{1}(0,x')$ and transformed by a linear system of
point spread function $h_{3}(x_{2},x')$, followed by a squarer,
viz. the bilinear transformation of Eq. (\ref{A2}).

On the other hand, using Eqs. (\ref{G2psi}), (\ref{I2x}), and
(\ref{basicEqObjInSignal}), the marginal coincidence rate,
measured when D$_{1}$ collects photons from all points in its
plane (i.e., acts as a bucket detector), is given by
\begin{eqnarray}\label{I2ObjInSig}
I^{(2)}(x_{2})&&=\int dx_{1}|\int
dx't(x')h_{1}(x_{1},x')h_{3}(x_{2},x')|^{2}\nonumber\\&&=\int\int
dx'dx''t^{*}(x')t(x'')g(x',x'')h^{*}_{3}(x_{2},x')h_{3}(x_{2},x'').
\end{eqnarray}
The quadratic transformation of the object $t(x)$ in Eq.
(\ref{I2ObjInSig}) is clearly the mathematical equivalent of the
bilinear transformation in Eq. (\ref{A1}), representing a
partially coherent imaging system. The function $g(x',x'')$ is
given by
\begin{equation}\label{gxxdash}
g(x',x'')=\int dx_{1} h_{1}^{*}(x_{1},x')h_{1}(x_{1},x''),
\end{equation}
and plays the role of the coherence function of the field.

Two limiting forms of $g(x',x'')$ are of interest. The first is
$g(x',x'')=\delta(x'-x'')$, which leads to
\begin{equation}
I^{(2)}(x_{2})=\int dx'|t(x')|^{2}|h_{3}(x_{2},x')|^{2},
\end{equation}
which is the equivalent of an incoherent imaging system [Eq.
(\ref{A3})]. The other limit is $g(x',x'')=f^{*}(x')f(x'')$, in
which case
\begin{equation}
I^{(2)}(x_{2})=|\int dx't(x')f(x')h_{3}(x_{2},x')|^{2},
\end{equation}
which is the equivalent of a coherent imaging system [Eq.
(\ref{A2})]. We can achieve the first limit by using a $2-f$
(Fourier transform) system or a $4-f$ (imaging) system for $h_{1}$
followed by a bucket detector. Moving the bucket detector in the
$2-f$ system away from the back focal plane or changing the area
of the detector would lead to a gradual transition from coherent
to incoherent imaging. This was performed experimentally in Ref.
[\cite{PR1}] where the change from coherent to incoherent imaging,
achieved by changing the detector aperture size in one beam, was
observed by monitoring the loss of the fringe visibility of a
double-slit placed in the other beam.

We now examine a few examples based on this configuration that
manifest its usefulness.

\subsubsection{Example: Fourier transform imaging}

Consider the system illustrated in Fig. \ref{object in the signal
beam examples 1 and 2}. The signal and idler systems are unfolded
for sake of clarity. We assume for simplicity in this and in the
following examples, except if indicated otherwise, that the pump
and the NLC are of infinite transverse extent. The signal arm
includes the object, $t$, and the system $h_{2}$ is nothing but
free space propagation at the signal wavelength $\lambda_{s}$ for
a distance $d_{s}$. The idler system comprises a lens of focal
length $f$ at a distance $d_{i}$ from the NLC, and a distance
$d_{2}$ from the detection plane, as shown in Fig. \ref{object in
the signal beam example 3}. Free-space propagation for the
distance $d_{s}$ at $\lambda_{s}$, and for the distance $d_{i}$ at
$\lambda_{i}$, may be substituted by free-space propagation for an
equivalent distance $d_{1}$ at wavelength $\lambda_{i}$, where
\begin{equation}\label{d1}
d_{1}=d_{i}+d_{s}\frac{\lambda_{s}}{\lambda_{i}}.
\end{equation}
If we take $d_{2}$ to be equal to the focal length of the lens
$f$, and also choose $d_{i}$ and $d_{s}$ such that $d_{1}=f$
according to Eq. (\ref{d1}), then the system becomes a Fourier
transform system with impulse response function
\begin{equation}\label{h3ObjInSigExample1}
h_{3}(x_{2},x')=\textmd{exp}(j2\pi[\frac{d_{s}}{\lambda_{s}}+\frac{d_{i}+f}{\lambda_{i}}])\textmd{exp}(-j\frac{2\pi}{\lambda_{i}f}x_{2}x').
\end{equation}

If we now take the illumination system $h_{1}$ to be uniform, so
that $h_{1}(0,x')=1$, then the overall system is a Fourier
transform system when the conditional coincidence rate is
considered. Equations (\ref{I20ObjInSig}) and
(\ref{h3ObjInSigExample1}) then yield
\begin{equation}
I_{0}^{(2)}(x_{2})=|T(\frac{2\pi}{\lambda_{i}f}x_{2})|^{2},
\end{equation}
where $T(q)$ is the Fourier transform of $t(x)$. The system simply
generates the diffraction pattern of the object distribution.

\subsubsection{Example: Ideal single-lens imaging}

In the same configuration depicted in Fig. \ref{object in the
signal beam examples 1 and 2}, we may choose the distance $d_{1}$,
calculated according to Eq. (\ref{d1}), and the distance $d_{2}$
to satisfy
\begin{equation}\label{IdealImagingEq}
\frac{1}{d_{1}}+\frac{1}{d_{2}}=\frac{1}{f},
\end{equation}
which is the geometrical-optics imaging equation of a thin lens of
focal length $f$. In this case the impulse response function of
the system $h_{3}$ becomes
\begin{equation}
h_{3}(x_{2},x')=\textmd{exp}(j2\pi[\frac{d_{s}}{\lambda_{s}}+\frac{d_{i}+d_{2}}{\lambda_{i}}])\textmd{exp}(j\frac{\pi
x_{2}^{2}}{\lambda_{i}d_{2}}[1-\frac{1}{M}])\delta(x_{2}-Mx'),
\end{equation}
where $M=-\frac{d_{2}}{d_{1}}$ is the magnification of the imaging
system. If the illumination system $h_{1}$ is uniform, then the
conditional coincidence rate $I_{0}^{(2)}(x_{2})$ is proportional
to the magnified object intensity transmittance,
$|t(\frac{x_{2}}{M})|^{2}$. The marginal coincidence rate
$I^{(2)}(x_{2})$, however, is
$I^{(2)}(x_{2})=g(\frac{x_{2}}{M},\frac{x_{2}}{M})|t(\frac{x_{2}}{M})|^{2}$,
where $g(x',x'')$ is given by Eq. (\ref{gxxdash}). If $g(x',x')$
is uniform over an area larger than that of the image,
$I^{(2)}(x_{2})$ becomes proportional to the magnified object
intensity transmittance as is the case for $I_{0}^{(2)}(x_{2})$.

Note that the lens may equivalently be put in the signal beam and
the distances readjusted so as to satisfy a condition similar to
Eq. (\ref{IdealImagingEq}). The system developed by Pittman {\it
et al.} \cite{TP1} is an example of this case in which the object
is placed directly in the plane of D$_{1}$ so that
$h_{1}(x_{1},x')=\delta(x_{1}-x')$.

\subsubsection{Example: Lens in the pump}

We now study another example where we manipulate the pump beam and
place the object in either the signal or idler paths. An example
of this configuration is the system proposed by Belinsky and
Klyshko \cite{AB3} and demonstrated experimentally by Pittman {\it
et al} \cite{TP2}.

The configuration is shown schematically in Fig. \ref{object in
the signal beam example 3}. A plane-wave pump beam is focused by a
lens of focal length $f$ , located at a distance $d<f$ from the
NLC. The pump wave front now has a radius of curvature $R=f-d$ at
the NLC entrance, and consequently acts as a lens or spherical
mirror in the advanced wave interpretation. The signal system is
comprised of free space propagation for a distance $d_{1}$ at
$\lambda_{s}$, followed by the object $t$ and an optical system
$h_{1}$. The idler system is simply free space propagation for a
distance $d_{2}$ at $\lambda_{i}$.

If the following relationship is satisfied by the various
distances and wavelengths:
\begin{equation}\label{IdealImagingEqLensInPump}
\frac{1}{\lambda_{s}d_{1}}+\frac{1}{\lambda_{i}d_{2}}=\frac{1}{\lambda_{p}R},
\end{equation}
then $h_{3}(x_{2},x')$, from Eq. (\ref{h3}), is
\begin{equation}
h_{3}(x_{2},x')=\textmd{exp}(j2\pi[\frac{d_{1}}{\lambda_{s}}+\frac{d_{2}}{\lambda_{i}}])\textmd{exp}(j\frac{\pi
x_{2}^{2}}{\lambda_{i}d_{2}}[1-\frac{1}{M}])\delta(x_{2}-Mx'),
\end{equation}
This is the impulse response function of an imaging system of
magnification $M=-\frac{d_{2}\lambda_{i}}{d_{1}\lambda_{s}}$, and
Eq. (\ref{IdealImagingEqLensInPump}) is the corresponding imaging
equation. Note the similarities between Eqs.
(\ref{IdealImagingEq}) and (\ref{IdealImagingEqLensInPump}),
despite the fact that the lens is in the signal beam for the
former and in the pump beam for the latter.

In the degenerate case where the signal and idler frequencies are
equal, $\lambda_{s}=\lambda_{i}=2\lambda_{p}$, this imaging
equation (Eq. (\ref{IdealImagingEqLensInPump})), becomes
\begin{equation}\label{IdealImagingEqLensInPumpDegenerate}
\frac{1}{d_{1}}+\frac{1}{d_{2}}=\frac{2}{R}.
\end{equation}
This is the imaging equation of a spherical mirror of radius of
curvature $R$, or is the geometrical-optics imaging equation of a
thin lens of focal length $\frac{R}{2}$. Both $I_{0}^{(2)}(x_{2})$
and $I^{(2)}(x_{2})$ are proportional to the magnified object
intensity transmittance $|t(\frac{x_{2}}{M})|^{2}$ if $h_{1}$ is
uniform.

\subsection{Object in Both Signal and Idler Beams}

If the signal and idler systems are identical and the object $t$
is placed at the same location in each, as shown in Fig.
\ref{object in both signal and idler beams}, then we may
substitute $h_{s}(x_{1},x)=h_{i}(x_{1},x)=\int
dx'h_{1}(x_{1},x')t(x')h_{2}(x',x)$ in Eq. (\ref{basicEq}) to
obtain
\begin{equation}\label{EpsiObjInBothSignalIdler}
\psi(x_{1},x_{2})=\int\int
dx'dx''t(x')t(x'')\psi_{c}(x',x'')h_{2}(x_{1},x')h_{2}(x_{2},x''),
\end{equation}
where
\begin{equation}\label{PsiC}
\psi_{c}(x',x'')=\int dxE_{p}(x)h_{1}(x',x)h_{1}(x'',x).
\end{equation}
Comparing Eq. (\ref{EpsiObjInBothSignalIdler}) with Eq. (\ref{A1})
shows that $\psi(x_{1},x_{2})$ is analogous to a partially
coherent imaging system, where $\psi_{c}(x',x'')$ plays the role
of the correlation function of the field. In this case, though, in
accordance with Eq. (\ref{G2psi}) the biphoton rate
$G^{(2)}(x_{1},x_{2})$ is a fourth-order nonlinear transformation
of $t$.

This system has been used \cite{AA3} to test the complementarity
of coherence and entanglement with the change of transverse size
of the pump beam, where $t$ was taken to be a double-slit. From
Eqs. (\ref{EpsiObjInBothSignalIdler}) and (\ref{PsiC}) it is clear
that for a small source the biphoton amplitude factorizes into a
function of $x_{1}$ and another function of $x_{2}$ (coherence),
while it is not factorizable (i.e., entangled) for a large
pump-beam size (entanglement) \cite{BS3}.

\subsection{Object in the Pump Beam}

In another imaging configuration, the object is placed in the pump
beam as illustrated in Fig. \ref{object in pump beam}. Equations
(\ref{G2psi}) and (\ref{basicEq}) give
\begin{equation}\label{G2x1x2ObjInPump}
G^{(2)}(x_{1},x_{2})=|\int dx
t(x)h_{s}(x_{1},x)h_{i}(x_{2},x)|^{2}
\end{equation}
provided that $E_{p}(x)$ is uniform over the object. Many
possibilities for imaging based on Eq. (\ref{G2x1x2ObjInPump}) can
be envisioned. For example, if both $h_{s}$ and $h_{i}$ are $2-f$
systems, the result is proportional to the squared magnitude of
the Fourier transform of $t$. In another example, if
$h_{s}(0,x)=1$, then
\begin{equation}\label{I02ObjInPump}
I_{0}^{(2)}(x_{2})=|\int dx t(x)h_{i}(x_{2},x)|^{2},
\end{equation}
and the behavior is that of a coherent imaging system. The object
is illuminated at the pump wavelength, while the measurement is
made at the much longer signal and idler wavelengths.

In a third example in which the signal and idler systems are
identical, and the coincidence is measured at the same position,
by use of a detector sensitive to the arrival of photon pairs (a
two-photon absorber, for example), then
\begin{equation}\label{G2x1x1ObjInPump}
G^{(2)}(x_{1},x_{1})=|\int dx t(x)h_{s}^{2}(x_{1},x)|^{2}.
\end{equation}
Again, the mathematical structure is that of a coherent imaging
system.

An interesting modification to this configuration would be to add
a $2-f$ system between the object and the crystal. In this case,
$E_{p}(x)=T(2\pi\frac{x}{\lambda_{p}f_{o}})$, where $f_{o}$ is the
focal length of the lens before the crystal. If the object is not
symmetric in $x$, then its Fourier transform is a complex
function. Yet the phase distribution of the object's spatial
spectrum is not lost, since the three-wave interaction process in
the NLC is coherent. If, in addition, we take both the signal and
idler configurations to be $2-f$ systems, the biphoton rate
becomes
\begin{equation}\label{G2x1x2ObjInPump2fBeforeNLC}
G^{(2)}(x_{1},x_{2})=|t([\frac{x_{1}}{\lambda_{s}f_{s}}+\frac{x_{2}}{\lambda_{i}f_{i}}]f_{o}\lambda_{p})|^{2}.
\end{equation}
where $f_{s}$ and $f_{i}$ are the focal lengths of the $2-f$
systems in the signal and idler beams, respectively. In this case,
in accordance with Eq. (\ref{I20x}), $I_{0}^{(2)}$ provides a
magnified image of the object $t$ with a magnification factor of
$\frac{\lambda_{i}f_{i}}{\lambda_{p}f_{o}}$.

\subsection{Object is the Detector}

In yet another imaging modality, illustrated in Fig. \ref{object
is the detector}, the object is a two-photon absorber; it is thus
a detector with quantum efficiency proportional to its absorbance
$t_{2}$. The biphoton rate, in this case $G^{(2)}(x_{1},x_{1})$,
is registered by some response of the object, such as emitted
photoelectrons or fluorescence \cite{MT1}. The signal/idler
optical system may, for example, be a single-lens imaging system
or a scanning system, as in scanning confocal microscopy
\cite{TW}. From Eqs. (\ref{G2psi}) and (\ref{basicEq})
\begin{equation}\label{G2x1x1ObjIsDetector}
G^{(2)}(x_{1},x_{1})=t_{2}(x_{1})S(x_{1}),
\end{equation}
where $S(x_{1})$ is an object illumination function. One choice
for $S$ would be a very narrow function , which would serve to
sample the object in the transverse plane. This could be achieved,
for example, by taking a pump of large transverse width and $2-f$
signal and idler systems. The size of the pump and the aperture of
the lens limit the transverse resolution.

There are other sensible choices for the illumination function
$S$. These can be implemented through either the pump profile, or
the system impulse response function, or both. The object $t_{2}$
would then be extracted by dividing the observed biphoton rate by
$S$. The object can also be uniformly illuminated by using a large
pump and a $4-f$ imaging system, in which case $S$ becomes almost
constant over a large portion of the object. We have studied this
system elsewhere and compared the longitudinal and transverse
resolutions to those of other schemes of microscopy that utilize
classical light\cite {MN}.

\subsection{Objects in Signal, Idler, and Pump Beams: Image Triple Correlation}

Because the biphoton optical system is based on three-wave mixing,
it inherently depends on three image distributions and therefore
offers a number of unique options for optical-image processing.
For example, if $4-f$ systems with aperture functions $t_{s}$ and
$t_{i}$ are placed in the signal and idler beams, respectively,
and a third object is placed in the pump beam such that the field
at the entrance to the NLC is $t_{p}$ then Eqs. (\ref{G2psi}) and
(\ref{basicEq}) yield
\begin{equation}\label{G2x1x2ObjInSigIdlerAndPump}
G^{(2)}(x_{1},x_{2})=|\int dx
t_{p}(x)T_{s}(\frac{2\pi}{\lambda_{s}f_{s}}(x-x_{1}))T_{i}(\frac{2\pi}{\lambda_{i}f_{i}}(x-x_{2}))|^{2},
\end{equation}
where $T_{s}(q)$ and $T_{i}(q)$ are the Fourier transforms of
$t_{s}(x)$ and $t_{i}(x)$; and $f_{s}$ and $f_{i}$ are the focal
lengths of the signal and idler $4-f$ systems, respectively. This
equation represents the magnitude of the triple correlation of the
three functions $t_{p}$, $T_{s}$, and $T_{i}$. Triple correlation
is useful in a number of signal-processing applications.  Of
course, if one of these three functions is uniform, the operation
becomes ordinary correlation.

One application for this configuration could be system
identification and coded-aperture imaging. In this application, a
linear, shift-invariant optical system is to be identified, i.e.,
its impulse response function measured. The system may be placed
in one of the two down-converted beam (say the signal) while a set
of $N$ known reference systems are placed, one at a time, in the
other beam (the idler) as the coincidence rate is measured. The
set of idler systems are also assumed to be linear,
shift-invariant with impulse response functions
\begin{equation}\label{LSIfunction}
h_{i}(x_{2},x)=h_{n}(x-x_{2}), \quad n=1, 2, ... N.
\end{equation}
Such systems may be generated by the use of a bank of apertures
(filters). Since the unknown system is shift-invariant, its
impulse response function is $h_{s}(x_{1},x)=h_{s}(x_{1}-x)$, so
that by virtue of Eqs. (\ref{G2psi}) and (\ref{basicEq}) the
biphoton rate measured at $x_{1}=x_{2}=0$ is given by
\begin{equation}\label{G2 00}
C_{n}=G^{(2)}(0,0)=|\int dx h_{s}(-x)h_{n}(x)|^{2},
\end{equation}
assuming the pump distribution to be uniform.

If $\{h_{n}(x)\}$ form a complete set of orthonormal functions,
then the measured coefficients $\{C_{n}\}$ are simply the squared
magnitudes of the coefficients of an expansion of the unknown
function $h_{s}(-x)$ in this basis. Under special conditions, the
phases can be retrieved, and the function $h_{s}(x)$ completely
reconstructed \cite{HS87}.

In the special case for which $h_{n}(x)=\delta(x+x_{n})$, so that
the idler field is sampled at positions $x_{n}$, Eq. (\ref{G2 00})
yields
\begin{equation}\label{}
C_{n}=|h_{s}(x_{n})|^{2},
\end{equation}
in which case the measured coincidence rates provide samples of
the magnitude of the impulse response function. A scanning system
can therefore be used to determine $|h_{s}(x)|$.

\section{RESOLUTION OF BIPHOTON IMAGING}

In all of the configurations studied in the previous section we
assumed a thin NLC and a narrow biphoton spectral bandwidth. Under
these assumptions the imaging resolution of all the configurations
is determined by the apertures placed in the system (including
those placed in the pump beam). When these apertures are not
accounted for, we obtain results reminiscent of classical {\it
geometric optics}, such as the imaging formulas in Eqs.
(\ref{IdealImagingEq}) and (\ref{IdealImagingEqLensInPump}). These
geometric-optics results are typical of the work that has been
carried out to date in entangled-photon imaging \cite{TP1,TP2}.

One of the advantages of our formalism is to facilitate deriving
the analog of {\it wave-optics} results for such systems when all
the physical parameters of the optical arrangement are accounted
for, using straightforward calculations similar to those of
classical wave optics \cite{BS1,MB}. In this section we examine
the effect of the various parameters of entangled-photon imaging
systems on the imaging resolution.

We take the width of the image formed by a point object
$t(x)=\delta(x)$ in the marginal coincidence rate [Eq.
(\ref{I2x})] as {\it a measure of the resolution of the
entangled-photon imaging system}. Another definition of resolution
may be based on the conditional coincidence rate [Eq.
(\ref{I20x})].

We begin by modifying the principal imaging equations [Eqs.
(\ref{G2psi}) and (\ref{basicEq})] by taking into consideration
the thickness of the NLC and the biphoton spectral bandwidth. We
assume a monochromatic pump beam of angular frequency $\omega_{p}$
and transverse distribution $E_{p}(x)$ at the entrance to a NLC of
thickness $\ell$. The coincidence rate
$G^{(2)}(x_{1},t_{1};x_{2},t_{2})$, with the detection times of
D$_{1}$ and D$_{2}$ now made explicit, is given by
\begin{equation}\label{G2x1t1x2t2psi}
G^{(2)}(x_{1},t_{1};x_{2},t_{2})=|\psi(x_{1},t_{1};x_{2},t_{2})|^{2}.
\end{equation}
Here $\psi(x_{1},t_{1};x_{2},t_{2})$ may be written in terms of a
biphoton spectral amplitude $\tilde{\psi}(x_{1},x_{2};\omega_{s})$
via
\begin{equation}\label{epsix1t1x2t2epsitilde}
\psi(x_{1},t_{1};x_{2},t_{2})=\textmd{exp}(-i\omega_{p}t_{1})\int_{\Omega}
d\omega_{s}\textmd{exp}(-i\omega_{s}(t_{1}-t_{2}))\tilde{\psi}(x_{1},x_{2};\omega_{s}),
\end{equation}
where $\Omega$ is the biphoton spectral bandwidth and
$\tilde{\psi}(x_{1},x_{2};\omega_{s})$ is given by \cite{BS3}
\begin{equation}\label{basicEqepsitilde}
\tilde{\psi}(x_{1},x_{2};\omega_{s})=\int\int
dq_{s}dq_{i}\Lambda(q_{s},q_{i};\omega_{s})H_{s}(x_{1},q_{s};\omega_{s})H_{i}(x_{2},q_{i};\omega_{p}-\omega_{s}),
\end{equation}
and the dispersion of the optical systems has been made explicit
in the signal and idler transfer functions $H_{s}$ and $H_{i}$,
which are Fourier transforms of the impulse response functions
$h_{s}$ and $h_{i}$ (with respect to the second argument),
respectively. The quantity $\Lambda(q_{s},q_{i};\omega_{s})$ in
Eq. (\ref{basicEqepsitilde}) is given by
\begin{equation}\label{LambdaFunction}
\Lambda(q_{s},q_{i};\omega_{s})=\tilde{E}_{p}(q_{s}+q_{i})\tilde{\xi}(q_{s},q_{i};\omega_{s});
\end{equation}
here $q$ is proportional to the transverse component of the
momentum vector (it is the spatial frequency in the transverse
plane), $\tilde{E}_{p}(q)$ is the Fourier transform of $E_{p}(x)$,
and $\tilde{\xi}(q_{s},q_{i};\omega_{s})$ is a phase-matching
function given by
\begin{equation}\label{ExitaTilde}
\tilde{\xi}(q_{s},q_{i};\omega_{s})=\ell
\textmd{sinc}(\frac{\ell}{2\pi}\Delta
r)\textmd{exp}(-j\frac{\ell}{2}\Delta r);
\end{equation}
and $\Delta
r=r_{p}(q_{s}+q_{i},\omega_{p})-r_{s}(q_{s},\omega_{s})-r_{i}(q_{i},\omega_{p}-\omega_{s})$;
$r_{j}(q,\omega)=\sqrt{n_{j}^{2}\frac{\omega^{2}}{c^{2}}-q^{2}}$,
$j=p, s, \textmd{and}\: i$, where $n_{j}$ is the NLC index of
refraction for the polarization and frequency of the $j^{th}$
field.

In most cases the detectors may be considered to be slow (i.e.,
their response time is large with respect to the inverse of the
bandwidth of the system, which is a reasonable assumption for
available photodetectors), and thus they measure a coincidence
rate that is averaged over a long time interval. The resulting
time averaged coincidence rate is \cite{BS3}
\begin{equation}\label{Cx1x2}
C(x_{1},x_{2})=\int_{\Omega}d\omega_{s}|\tilde{\psi}(x_{1},x_{2};\omega_{s})|^{2},
\end{equation}
showing that the time averaged coincidence rate for a slow
detector is an incoherent sum of the biphoton spectral amplitudes
over the bandwidth of the system. The spectrum of the
down-converted biphotons can be quite large and the dispersion of
the optical components must be considered carefully just as
dispersion must be in ultrafast pulsed optics.

We also define conditional and marginal time-averaged coincidence
rates as
\begin{equation}\label{Cx2}
C(x_{2})=\int dx_{1}C(x_{1},x_{2}),
\end{equation}
\begin{equation}\label{C0x2}
C_{0}(x_{2})=C(0,x_{2}),
\end{equation}
respectively. It is obvious that when only a narrow spectral
bandwidth is considered, $C(x_{2})$ and $C_{0}(x_{2})$ coincide
with $I(x_{2})$ and $I_{0}(x_{2})$, respectively.

We now proceed to study the resolution of a representative
configuration considered in Section 2: {\it object in the signal
(or idler) beam}. The biphoton spectral amplitude of this system,
illustrated in Fig. \ref{object in the signal beam}, now taking
into consideration the thickness of the crystal and spectral
bandwidth of the system, is given by
\begin{equation}\label{EpsiTildeObjInSig}
\tilde{\psi}(x_{1},x_{2};\omega_{s})=\int
dx't(x')h_{1}(x_{1},x';\omega_{s})h_{3}(x_{2},x';\omega_{s}),
\end{equation}
where
\begin{equation}\label{h3EpsiTilde}
h_{3}(x_{2},x';\omega_{s})=\int\int
dq_{s}dq_{i}\Lambda(q_{s},q_{i};\omega_{s})H_{2}(x',q_{s};\omega_{s})H_{i}(x_{2},q_{i};\omega_{p}-\omega_{s}).
\end{equation}
We assume throughout that the object is thin and non-dispersive.
To determine the resolution of this imaging configuration we take
$t(x)=\delta(x)$, whereupon Eq.(\ref{EpsiTildeObjInSig}) becomes
$\tilde{\psi}(x_{1},x_{2};\omega_{s})=h_{1}(x_{1},0;\omega_{s})h_{3}(x_{2},0;\omega_{s})$,
and consequently
\begin{equation}\label{Cx2BasicObjInSig}
C(x_{2})=\int_{\Omega}d\omega_{s}|h_{3}(x_{2},0;\omega_{s})|^{2}g(\omega_{s}),
\end{equation}
\begin{equation}\label{C0x2BasicObjInSig}
C_{0}(x_{2})=\int_{\Omega}d\omega_{s}|h_{3}(x_{2},0;\omega_{s})|^{2}g_{o}(\omega_{s}),
\end{equation}
where $g(\omega)=\int dx|h_{1}(x,0;\omega)|^{2}$ and
$g_{o}(\omega)=|h_{1}(0,0;\omega)|^{2}$. Note that the system
$h_{1}$ affects the imaging resolution only through introducing an
effective spectral bandwidth that may be ignored if it is larger
than that of $h_{3}$.

As a concrete example we consider the system examined in Section
2.A.2, which is the second example of {\it object in the signal
beam} configuration, namely {\it ideal single-lens imaging},
illustrated in Fig. \ref{object in the signal beam examples 1 and
2}. We assume, at first, a plane wave pump, so that
$h_{3}(x_{2},0;\omega_{s})$ simplifies to
\begin{equation}\label{h30EpsiTildePlaneWavePump}
h_{3}(x_{2},0;\omega_{s})=\int
dq_{s}\tilde{\xi}(q_{s},-q_{s};\omega_{s})H_{2}(0,q_{s};\omega_{s})H_{i}(x_{2},-q_{s};\omega_{p}-\omega_{s}).
\end{equation}
In this example, the transfer functions of the systems $h_{2}$ and
$h_{i}$ are given by
\begin{equation}\label{H2}
H_{2}(0,q_{s};\omega_{s})=\textmd{exp}(jk_{s}d_{s})\textmd{exp}(-j\frac{d_{s}q_{s}^{2}}{2k_{s}}),
\end{equation}
\begin{equation}\label{Hi}
H_{i}(x_{2},-q_{s};\omega_{p}-\omega_{s})=
\textmd{exp}(jk_{i}(d_{1}+d_{2}))
\textmd{exp}(j\frac{k_{i}x_{2}^{2}}{2d_{2}})
\textmd{exp}(-j\frac{d_{i}q_{s}^{2}}{2k_{i}})
P_{g}(q_{s}+\frac{k_{i}x_{2}}{d_{2}}),
\end{equation}
where $P_{g}(q)$ is the Fourier transform of
$p(x)\textmd{exp}(j\frac{k_{i}x^{2}}{2}[\frac{1}{d_{2}}-\frac{1}{f}])$
with respect to $x$, and $p(x)$ is the lens aperture. Substituting
Eqs. (\ref{ExitaTilde}), (\ref{H2}) and (\ref{Hi}) into Eq.
(\ref{h30EpsiTildePlaneWavePump}) we obtain
$h_{3}(x_{2},0;\omega_{s})$ which we then use in Eqs.
(\ref{Cx2BasicObjInSig}) and (\ref{C0x2BasicObjInSig}) to estimate
the resolution.

There are two techniques to implement this system in an actual
setup. In one technique the NLC is adjusted for non-collinear
SPDC, and one beam (usually chosen by a pinhole) is directed into
the system $h_{s}$ and the other into $h_{i}$. Another technique
is to adjust the NLC for collinear SPDC and then separate the two
photons comprising the biphoton. In type-II SPDC (where the signal
and idler photons have orthogonal polarizations) one can use a
polarizing beam splitter to separate the biphoton. On the other
hand, in type-I SPDC (where the signal and idler photons have the
same polarization) the use of a non-polarizing beam-splitter will
separate the pair into the two output ports of the beam splitter
in 50$\%$ of the trials, and send the pair together into one
output port in the rest of the trials. In the latter case, the
trials do not contribute to the coincidence measurements carried
out by the detectors D$_{1}$ and D$_{2}$ together with the
coincidence detection circuit, and thus may be ignored.

Assuming a thin NLC, narrow spectral bandwidth, a plane-wave pump,
and degenerate collinear down-conversion (where the signal and
idler photons are separated using the method outlined above), one
obtains the familiar diffraction pattern of a diffraction-limited
imaging system. For a rectangular lens aperture of width $D$ and
focal length $f$ the result is
$C(x_{2})\propto|\textmd{sinc}(\frac{x_{2}}{2\lambda_{o}F_{\#}})|^{2}$,
where $F_{\#}=\frac{f}{D}$ is the the F-number of the lens and
$\lambda_{o}=2\lambda_{p}$ is the wavelength of the degenerate
down-converted photons. This is the best one can obtain; we
demonstrate in the following that relaxing any of the restrictions
indicated above will degrade the resolution.

Our calculations have been carried out using a $\beta$-barium
borate (BBO)  NLC that is illuminated with a pump of wavelength
$\lambda_{p}=325$ nm (which corresponds to the ultraviolet line of
a He-Cd laser), with a cut-angle of $36.44^{\circ}$ that
corresponds to degenerate collinear type-I SPDC. Increasing the
cut-angle beyond $36.44^{\circ}$ yields non-collinear degenerate
SPDC while decreasing the cut-angle below this value yields
collinear non-degenerate SPDC \cite{AA1}.

We first consider the effect of the finite thickness of the NLC.
One effect is that the distance $d_{1}$, used in the imaging
formula presented in Eq. (\ref{IdealImagingEq}), is modified to
become
\begin{equation}\label{modified d1}
d_{1}=d_{i}+d_{s}\frac{\lambda_{s}}{\lambda_{i}}+\ell_{eq},
\end{equation}
in contrast to that given in Eq. (\ref{d1}). The quantity
$\ell_{eq}$ is an equivalent length for the NLC that is related to
the physical length $\ell$ by
\begin{equation}\label{lequivalent}
\ell_{eq}=\frac{\ell}{2\lambda_{i}}(\frac{\lambda_{s}}{n_{s}}+\frac{\lambda_{i}}{n_{i}}).
\end{equation}
For the degenerate case ($\lambda_{s}=\lambda_{i}=\lambda_{o}$)
this expression simplifies to $\ell_{eq}=\frac{\ell}{n_{o}}$,
where $n_{o}$ is the index of refraction of the NLC at the
degenerate wavelength. In other words, the thickness of the NLC
must be accounted for in calculating the distances in the
experimental arrangement in order to satisfy the imaging formula
of Eq. (\ref{IdealImagingEq}).

Figure \ref{effect of NLC thickness} shows $C(x_{2})$ for several
values of $\ell$, assuming degenerate collinear down-conversion
and narrow spectral bandwidth. The distances in the configuration
are chosen such that $d_{1}=d_{2}=2f$ (taking into account the
effect of $\ell$ on $d_{1}$ according to Eq. (\ref{modified d1})),
which corresponds to an imaging system of unity magnification. For
$\ell=0.1$ mm one obtains the diffraction limited distribution
corresponding to the thin NLC case. When $\ell$ increases the
distribution widens, signifying a loss of imaging resolution, as
is evident for the $\ell=1$-mm and $\ell=10$-mm curves. This
result may be easily understood when one considers the fact that
the NLC acts as a spatial filter through the phase-matching
function $\tilde{\xi}(q_{s},q_{i};\omega_{s})$ defined in Eq.
(\ref{ExitaTilde}). The collinear SPDC case corresponds to a
low-pass spatial filter with a cut-off frequency that is inversely
proportional to $\ell$ and hence the resolution degrades as the
NLC thickness increases.

The spectral bandwidth of the system has a similar effect on the
imaging resolution, which decreases with increased bandwidth.
Figure \ref{effect of bandwidth} shows $C(x_{2})$ for several
values of $\rho=\frac{\Omega}{\omega_{p}}$. These plots were
obtained for a NLC of thickness $\ell=1$ mm, collinear SPDC, and a
plane-wave pump. According to Eq. (\ref{modified d1}) $d_{1}$ is a
function of wavelength (and so is $\ell_{eq}$ via Eq.
(\ref{lequivalent})), so that only one pair of signal/idler
wavelengths satisfy the imaging formula in Eq.
(\ref{IdealImagingEq}). All biphotons with other signal/idler pair
wavelengths are defocused, and hence their contribution to
$C(x_{2})$ leads to a reduction in resolution. The plots in Fig.
\ref{effect of bandwidth} were obtained assuming that Eq.
(\ref{IdealImagingEq}) is satisfied by the degenerate signal/idler
wavelengths, and that at these wavelengths $d_{1}=d_{2}=2f$.

Finally, the finite transverse width of the pump field also
degrades the resolution. This can be understood by noting that
smaller pump size reduces entanglement of the signal and idler
photons \cite{BS3}. As a result, the quantum state of the light
emitted by the NLC becomes separable and thus $C(x_{2})$ and
$C_{o}(x_{2})$ simply become the intensity of the idler beam
(which depends on $h_{i}$), but are independent of the signal beam
\cite{AA2}. No information about the system $h_{s}$, which
includes the object to be imaged in this case, may be extracted
from the measurements carried out in the idler beam.

Figure \ref{effect of pump width} shows plots of $C(x_{2})$ for
various values of the transverse width of the pump, denoted $B$.
The calculations were performed taking $\ell=1$ mm, assuming
collinear degenerate SPDC, and the presence of narrowband spectral
filters in the system. Distances were chosen such that
$d_{1}=d_{2}=2f$.

\section{CONCLUSION}

We have presented a Fourier-optics analysis of various imaging
configurations using the unique features of spontaneous parametric
down-conversion (SPDC) as a two-photon source. SPDC is a
three-wave mixing process; the pump, signal, and idler are coupled
through the phase-matching conditions. We investigated several
imaging and image-processing configurations that utilize the
quantum correlations among these three fields. Our formalism was
also used to study the resolution of these entangled-photon
imaging configurations.

\section*{APPENDIX: THE OPTICAL BILINEAR TRANSFORMATION}

\setcounter{equation}{0}
\renewcommand{\theequation}
      {\mbox{{A}{\arabic{equation}}}}

We present a brief overview of the theory of classical imaging in
the framework of the bilinear optical transformation. The
equations are formulated in such a way so as to facilitate
convenient comparisons with the two-photon and biphoton cases
presented in the text.

Because of the quadratic relation between the optical field and
the optical intensity, imaging systems are typically described by
a bilinear transformation \cite{BS4}. A general bilinear
transformation is expressed as
\begin{equation}\label{A1}
g(x_{1})=\int\int dx'dx''f^{*}(x')f(x'')q(x_{1};x',x''),
\end{equation}
where $f(x)$ is the input function, $q(x_{1};x',x'')$ is the
double impulse response function (DIR), and $g(x)$ is the output
function. In general $f(x)$ is complex, but $g(x)$ is guaranteed
to be real when the symmetry condition
$q(x_{1};x',x'')=q^{*}(x_{1};x'',x')$ is satisfied. The DIR
completely characterizes the bilinear system. This transformation
represents, in general, the imaging system depicted in Fig.
\ref{bilinear optics}. The input function $f(x)$ represents the
transparency $t(x)$; the DIR is a combination of the second-order
correlation function of the illumination $G^{(1)}(x',x'')$ and the
impulse response function $h(x_{1},x)$ of the linear optical
system; and the output $g(x)$ represents the intensity measured by
the optical detector.

In the ideal case
$q(x_{1};x',x'')=\delta(x_{1}-x')\delta(x_{1}-x'')$, whereupon
$g(x_{1})=|t(x_{1})|^{2}$, so that the system is a squarer with
zero spread. When the DIR factorizes in the form
$q(x_{1};x',x'')=h^{*}(x_{1},x')h(x_{1},x'')$, the output of the
system is given by
\begin{equation}\label{A2}
g(x_{1})=|\int dx'f(x')h(x_{1},x')|^{2}.
\end{equation}
Equation (\ref{A2}) is easily recognized as the output intensity
of a coherent imaging system with impulse response function
$h(x_{1},x')$ and an input complex field $f(x')$. When the DIR
takes the form
$q(x_{1};x',x'')=h^{*}(x_{1},x')h(x_{1},x'')\delta(x'-x'')$, we
obtain
\begin{equation}\label{A3}
g(x_{1})=\int dx'|f(x')|^{2}|h(x_{1},x')|^{2},
\end{equation}
which is the output of an incoherent system with point spread
function $|h(x_{1},x')|^{2}$ and input intensity $|f(x')|^{2}$. In
general, partially coherent imaging can be represented by a
bilinear system with a DIR given by
$q(x_{1};x',x'')=\gamma(x',x'')h^{*}(x_{1},x')h(x_{1},x'')$, where
$\gamma(x',x'')$ represents the correlation function of the input
light, and $h(x_{1},x')$ is the coherent impulse response
function. When $\gamma(x',x'')=1$, we recover coherent imaging
whereas when $\gamma(x',x'')=\delta(x'-x'')$, we recover
incoherent imaging.

Entangled-photon imaging, like partially coherent imaging, is
described by a bilinear system, with partial entanglement assuming
the role of partial coherence \cite{BS3}.

\section*{ACKNOWLEDGMENTS}

This work was supported by the National Science Foundation; by the
Center for Subsurface Sensing and Imaging Systems (CenSSIS), an
NSF engineering research center; and by the David $\&$ Lucile
Packard Foundation.

B. E. A. Saleh's e-mail address is besaleh@bu.edu.

\newpage
\begin{figure}[hb]
\caption{Biphoton imaging using photon pairs generated by
spontaneous parametric down-conversion. NLC stands for nonlinear
crystal; D$_{1}$ and D$_{2}$ are single-photon detectors at
locations $x_{1}$ and $x_{2}$, respectively;
$G^{(2)}(x_{1},x_{2})$ is the biphoton rate; $h_{p}(x,x')$,
$h_{s}(x_{1},x)$, and $h_{i}(x_{2},x)$ are the impulse response
functions of the optical systems placed in the paths of the pump,
signal, and idler beams, respectively.} \label{overall layout}
\end{figure}

\begin{figure}[hb]
\caption{{\it Object in the signal beam} configuration. $E_{p}$ is
the pump field at the entrance to the NLC; $h_{1}(x_{1},x')$ and
$h_{2}(x',x)$ are the impulse response functions of the optical
systems placed in the signal beam; $h_{i}(x_{2},x)$ is the impulse
response function of the optical system placed in the idler beam;
$t(x')$ is the object to be imaged, placed in the signal beam.}
\label{object in the signal beam}
\end{figure}

\begin{figure}[hb]
\caption{{\it Object in the signal beam} configuration of examples
1 and 2 displayed in an unfolded picture. $E_{p}$,
$h_{1}(x_{1},x')$, and $t(x')$ are the same as in Fig. \ref{object
in the signal beam}; $f$ is the focal length of a lens placed in
the idler beam. See text for details.} \label{object in the signal
beam examples 1 and 2}
\end{figure}

\begin{figure}[hb]
\caption{{\it Object in the signal beam} configuration of example
3 displayed in an unfolded mode. $h_{1}(x_{1},x')$, and $t(x')$
are the same as in Fig. \ref{object in the signal beam}; a lens is
placed in the pump beam and is represented here by the dotted lens
of focal length $f$. See text for details.} \label{object in the
signal beam example 3}
\end{figure}

\begin{figure}[hb]
\caption{{\it Object in both signal and idler beams}
configuration. $h_{1}(x_{1},x')$, $h_{2}(x',x)$ are the impulse
response functions of the optical systems placed in the path of
both the signal and idler beams; and $t(x')$ is the object to be
imaged.} \label{object in both signal and idler beams}
\end{figure}

\begin{figure}[hb]
\caption{{\it Object in the pump beam} configuration.
$h_{s}(x_{1},x)$, $h_{i}(x_{2},x)$ are as in Fig. \ref{overall
layout}; and $t(x)$ is the object to be imaged placed in the pump
beam .} \label{object in pump beam}
\end{figure}

\begin{figure}[hb]
\caption{{\it Object is the detector} configuration.
$h_{1}(x_{1},x)$, is the impulse response of the optical system
placed in the signal and idler paths; and D is a two-photon
detector at location $x_{1}$.} \label{object is the detector}
\end{figure}

\begin{figure}[hb]
\caption{Configuration for triple correlation. $h_{2s}(x',x)$ and
$h_{1s}(x_{1},x')$ are the impulse response functions of the
optical systems placed in the signal beam; $h_{2i}(x'',x)$ and
$h_{1i}(x_{2},x'')$ are the impulse response functions of the
optical systems placed in the idler beam; $t_{p}(x)$, $t_{s}(x')$,
and $t_{i}(x'')$ are the three objects to be correlated.}
\label{triple correlation}
\end{figure}

\begin{figure}[hb]
\caption{Effect of NLC thickness $\ell$ on the imaging resolution
of {\it object in the signal beam} configuration. Plots of
normalized time-averaged marginal coincidence rate $C(x_{2})$
versus detector D$_{2}$ location $x_{2}$ normalized with respect
to $x_{c}=2\lambda_{o}F_{\#}$, for $\ell=0.1, 1, \textmd{and}\:
10$ mm; $\lambda_{o}=650$ nm, $F_{\#}=5$.}\label{effect of NLC
thickness}
\end{figure}

\begin{figure}[hb]
\caption{Effect of the biphoton bandwidth $\Omega$ on the imaging
resolution of {\it object in the signal beam} configuration. Plots
of normalized $C(x_{2})$ versus $x_{2}$ normalized with respect to
$x_{c}=2\lambda_{o}F_{\#}$ are shown for
$\rho=\frac{\Omega}{\omega_{p}}=0.001, 0.01, \textmd{and}\: 0.02$;
$\lambda_{o}=650$ nm, $F_{\#}=5$. The NLC is adjusted for
collinear SPDC and is of thickness 1 mm.}\label{effect of
bandwidth}
\end{figure}

\begin{figure}[hb]
\caption{Effect of the transverse width of the pump $B$ on the
imaging resolution of {\it object in the signal beam}
configuration. Plots of normalized $C(x_{2})$ versus $x_{2}$
normalized with respect to $x_{c}=2\lambda_{o}F_{\#}$ are shown
for $B=2, 1, 0.5, \textmd{and} \: 0.1$ mm; $\lambda_{o}=650$ nm,
$F_{\#}=5$. }\label{effect of pump width}
\end{figure}

\begin{figure}[hb]
\caption{Classical partially-coherent imaging. The quantity
$G^{(1)}(x',x'')$ is the second-order correlation function of the
optical field; $t$ is the object to be imaged; $h(x_{1},x)$ is the
impulse response function of the imaging system; D is a detector
placed at position $x_{1}$ that records the intensity $I(x_{1})$.
}\label{bilinear optics}
\end{figure}

\newpage

\begin{figure}[h]
 \epsfxsize=6.8in \epsfysize=5.2in
 \epsffile{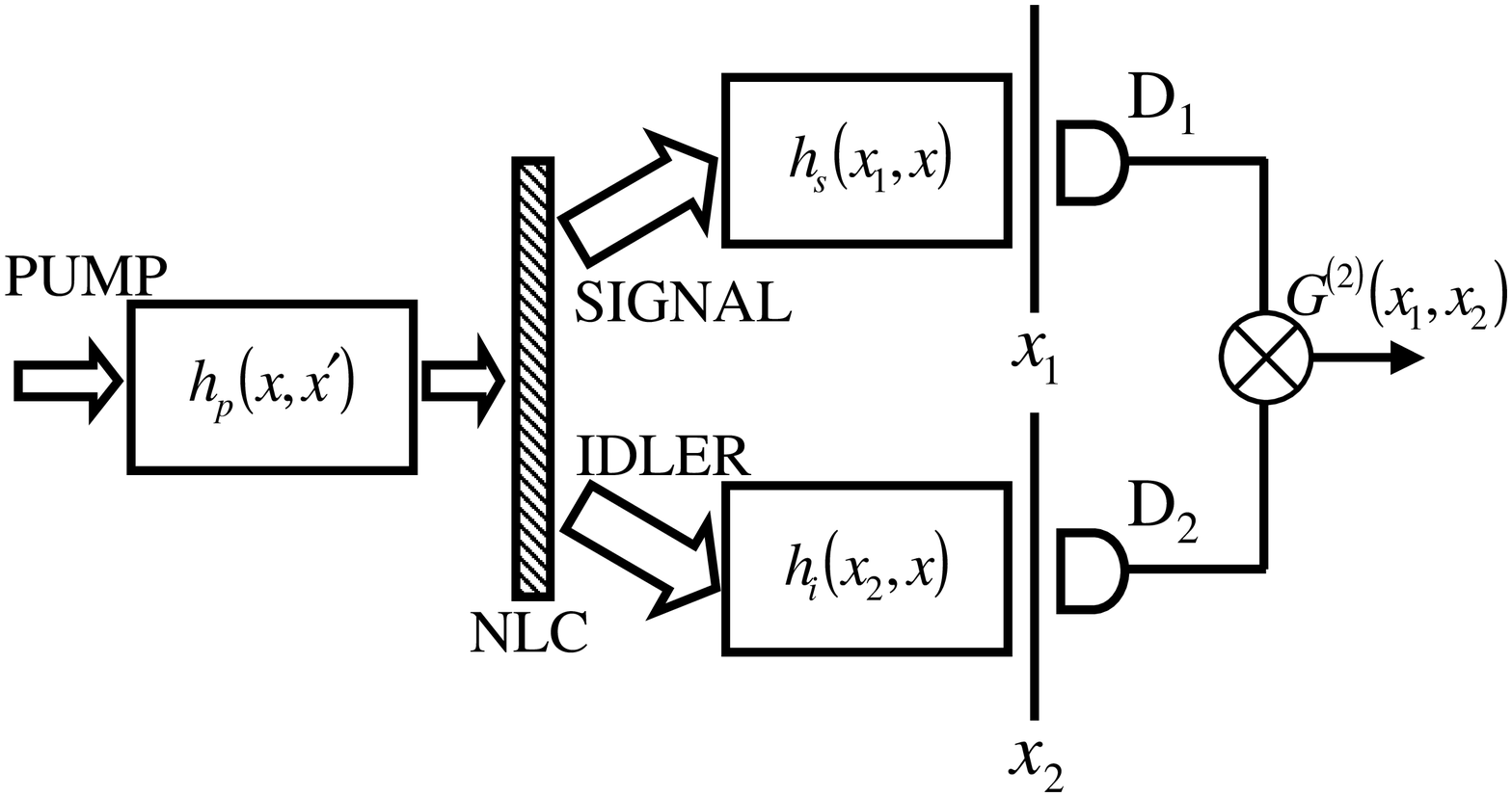}
 \end{figure}
\vskip6cm Figure 1, A. F. Abouraddy

\begin{figure}[h]
 \epsfxsize=6.8in \epsfysize=5.2in
 \epsffile{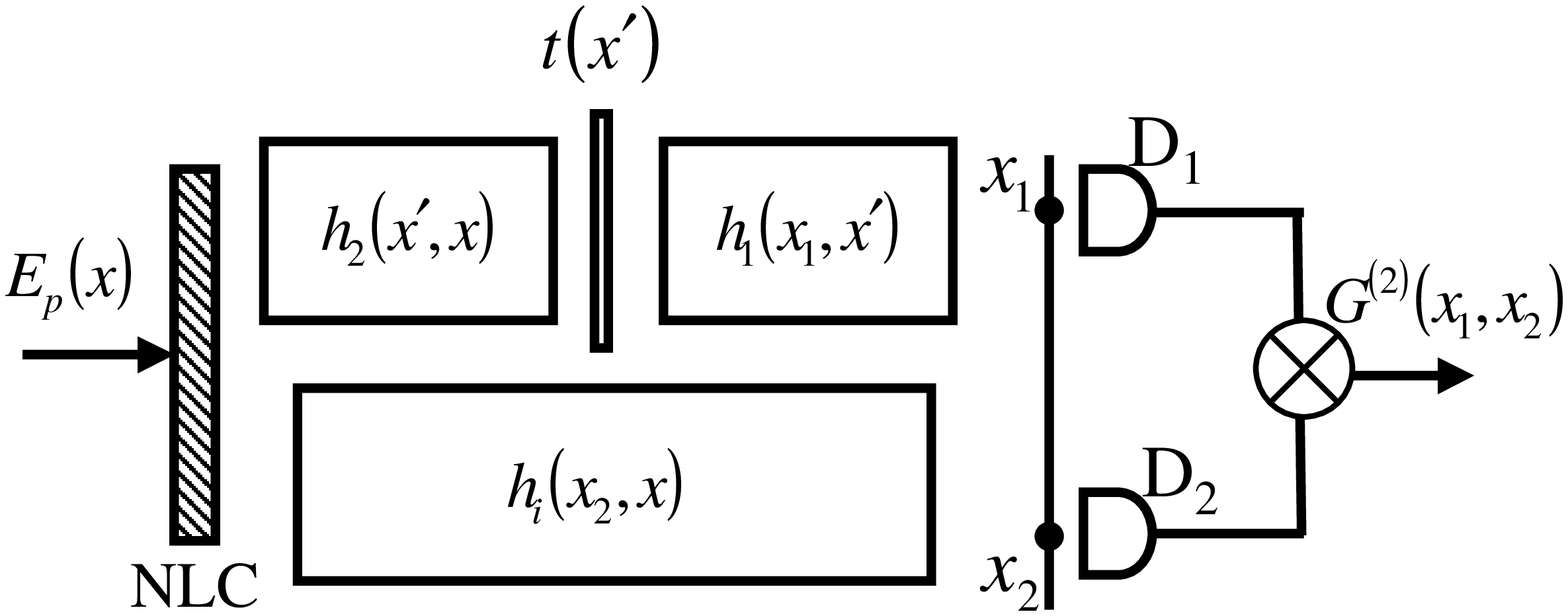}
 \end{figure}
\vskip6cm Figure 2, A. F. Abouraddy

\begin{figure}[h]
 \epsfxsize=6.8in \epsfysize=5.2in
 \epsffile{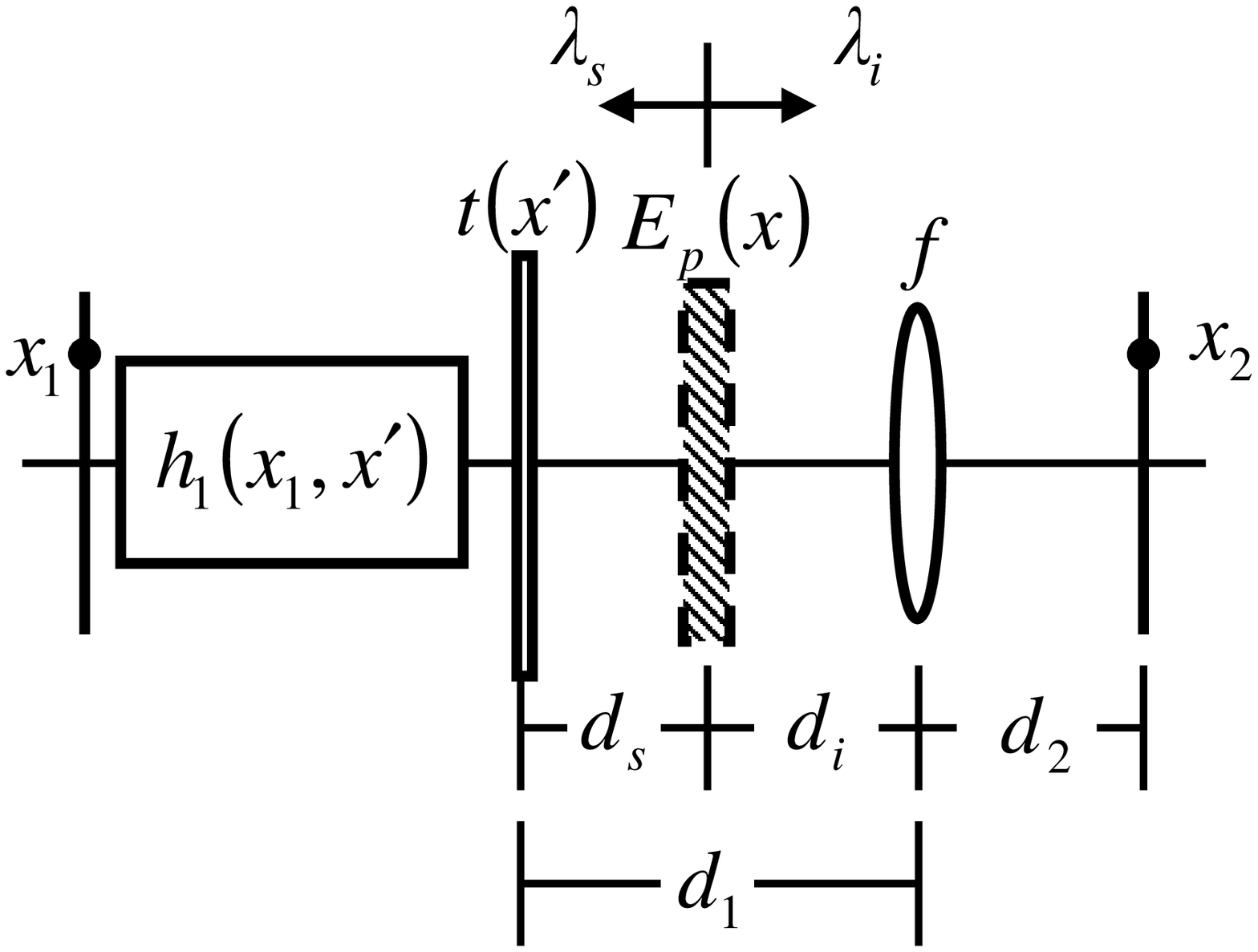}
 \end{figure}
\vskip6cm Figure 3, A. F. Abouraddy

\begin{figure}[h]
 \epsfxsize=6.8in \epsfysize=5.2in
 \epsffile{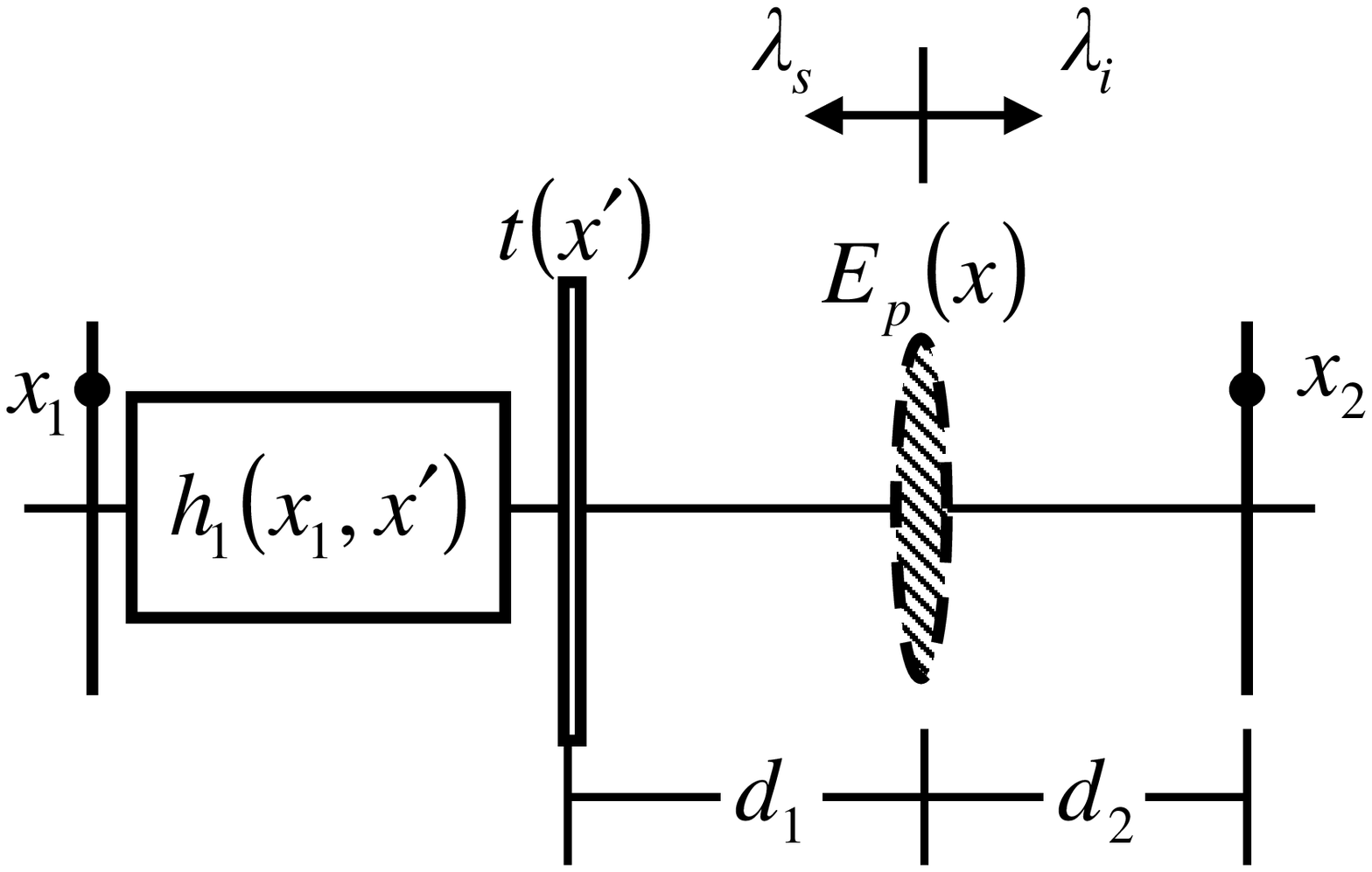}
 \end{figure}
\vskip6cm Figure 4, A. F. Abouraddy

\begin{figure}[h]
 \epsfxsize=6.8in \epsfysize=5.2in
 \epsffile{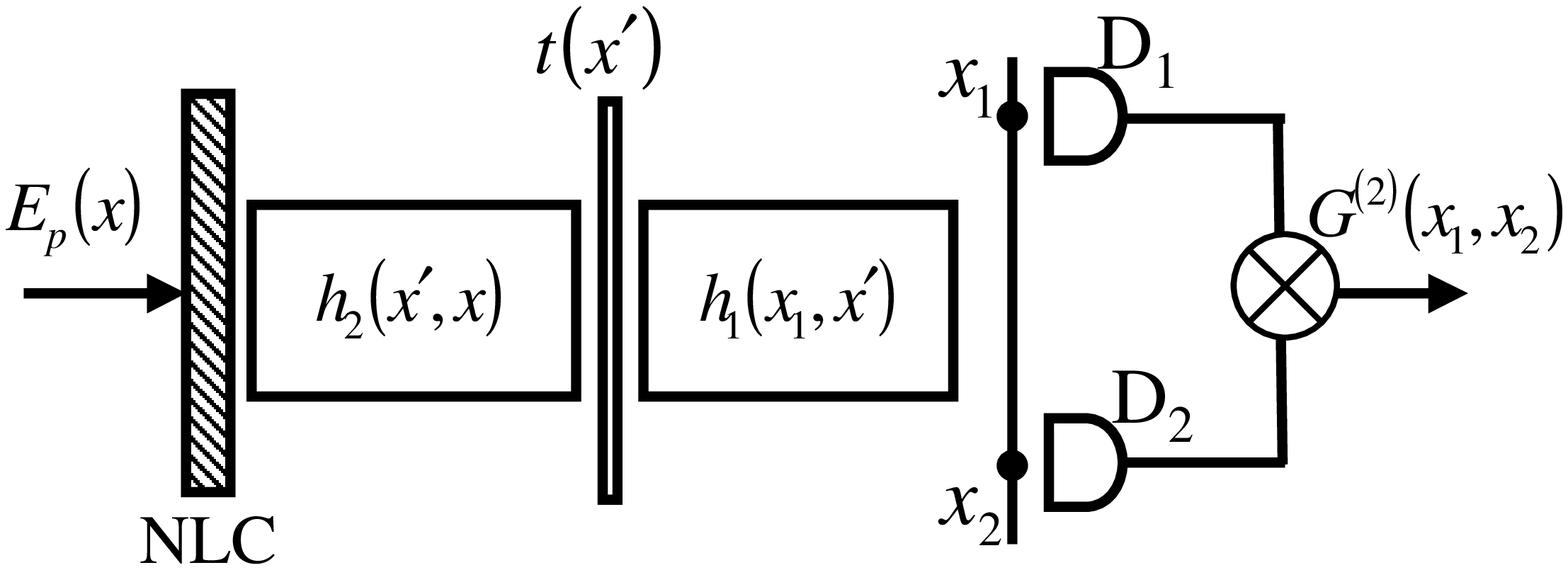}
 \end{figure}
\vskip6cm Figure 5, A. F. Abouraddy

\begin{figure}[h]
 \epsfxsize=6.8in \epsfysize=5.2in
 \epsffile{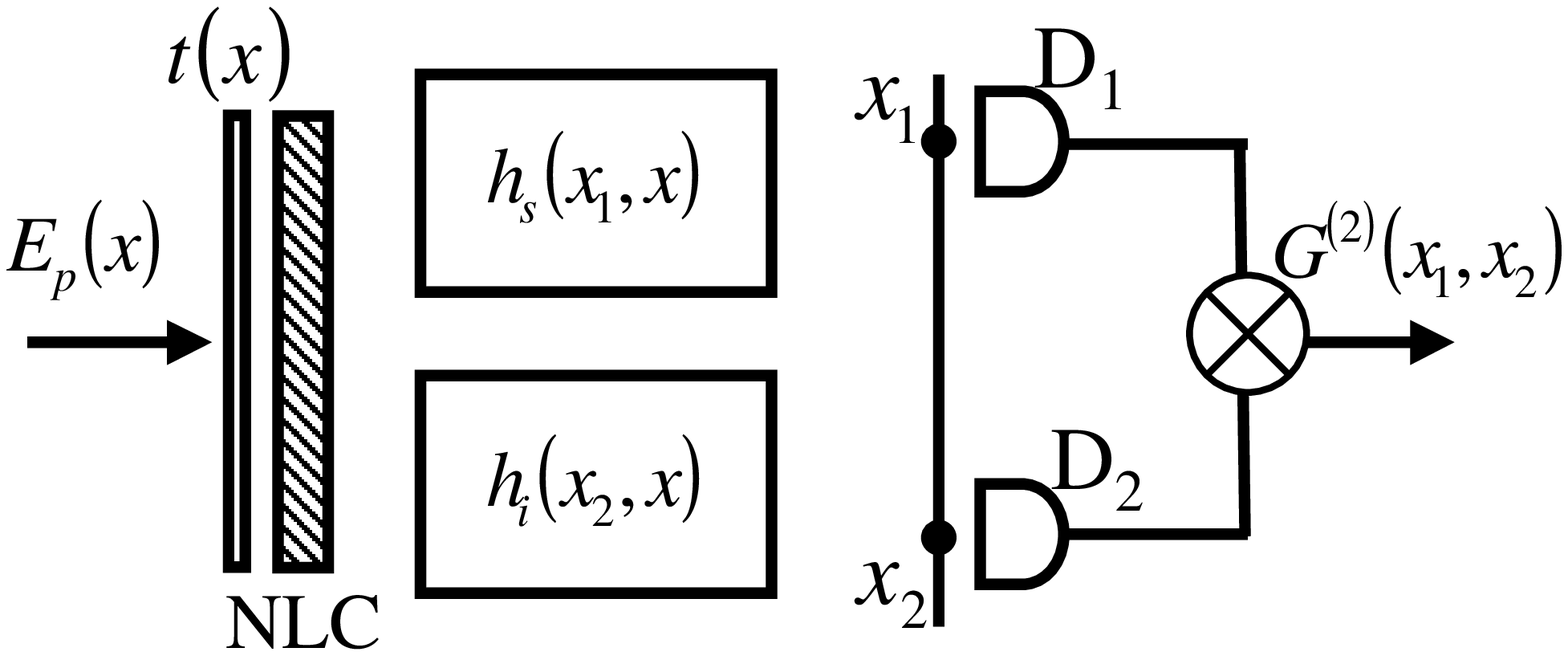}
 \end{figure}
\vskip6cm Figure 6, A. F. Abouraddy

\begin{figure}[h]
 \epsfxsize=6.8in \epsfysize=5.2in
 \epsffile{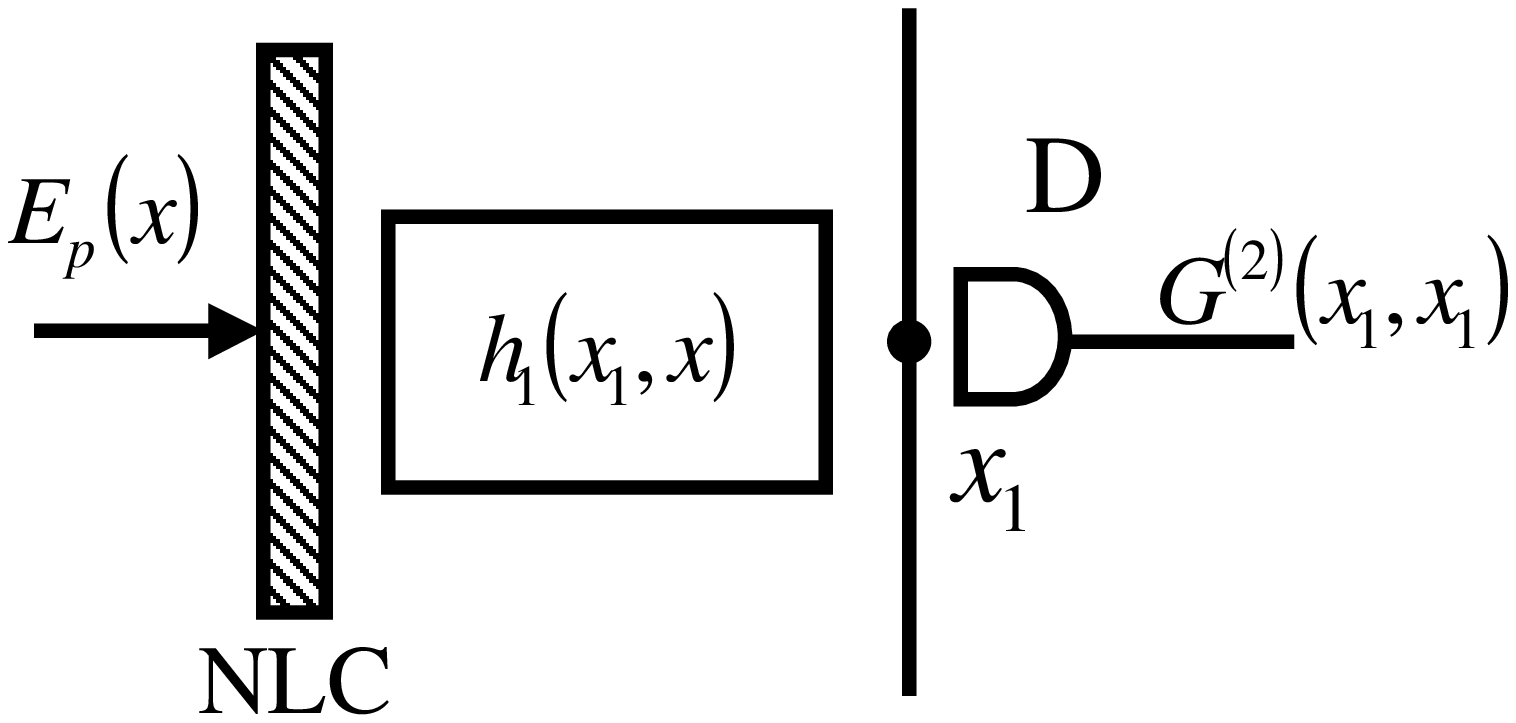}
 \end{figure}
\vskip6cm Figure 7, A. F. Abouraddy

\begin{figure}[h]
 \epsfxsize=6.8in \epsfysize=5.2in
 \epsffile{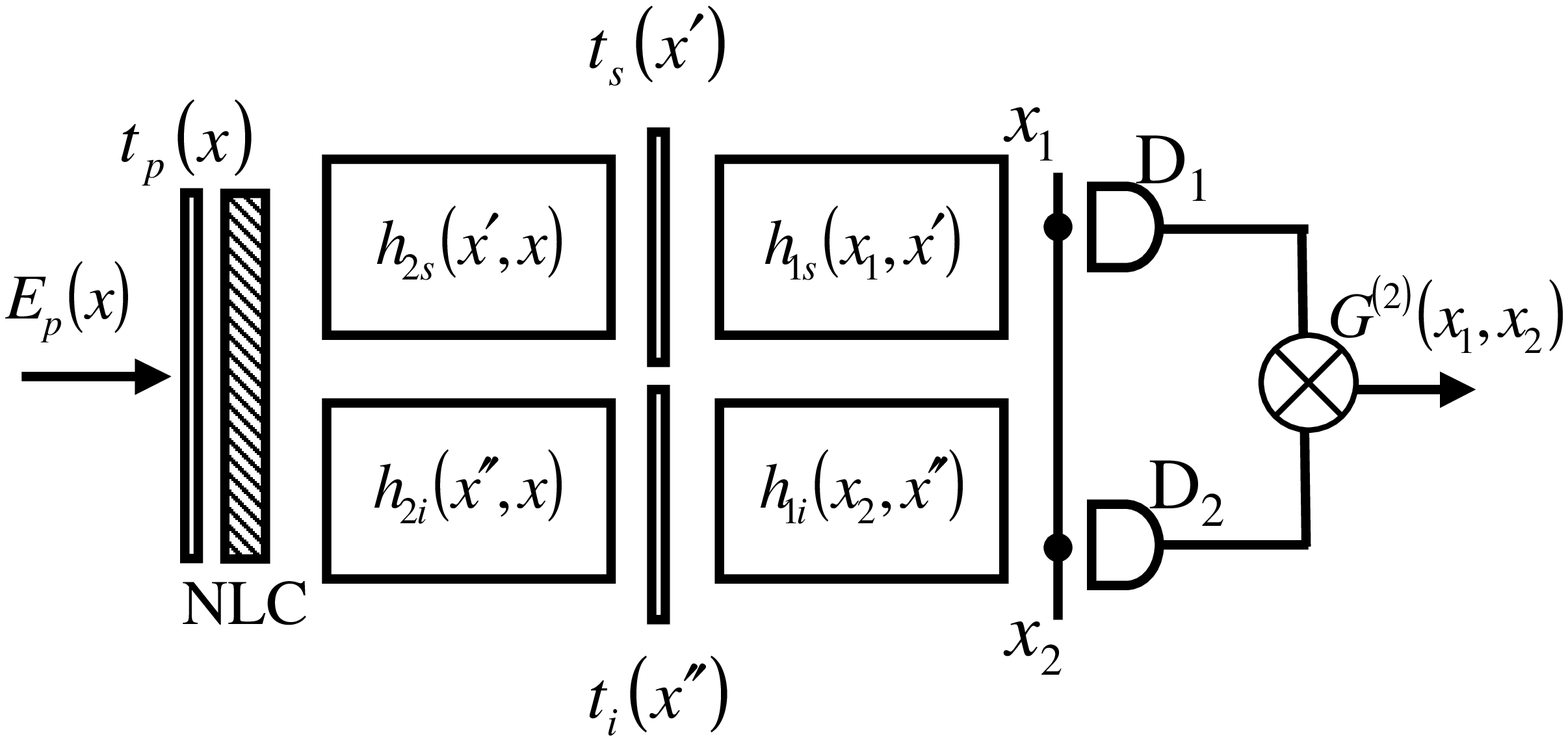}
 \end{figure}
\vskip6cm Figure 8, A. F. Abouraddy

\begin{figure}[h]
 \epsfxsize=6.8in \epsfysize=5.2in
 \epsffile{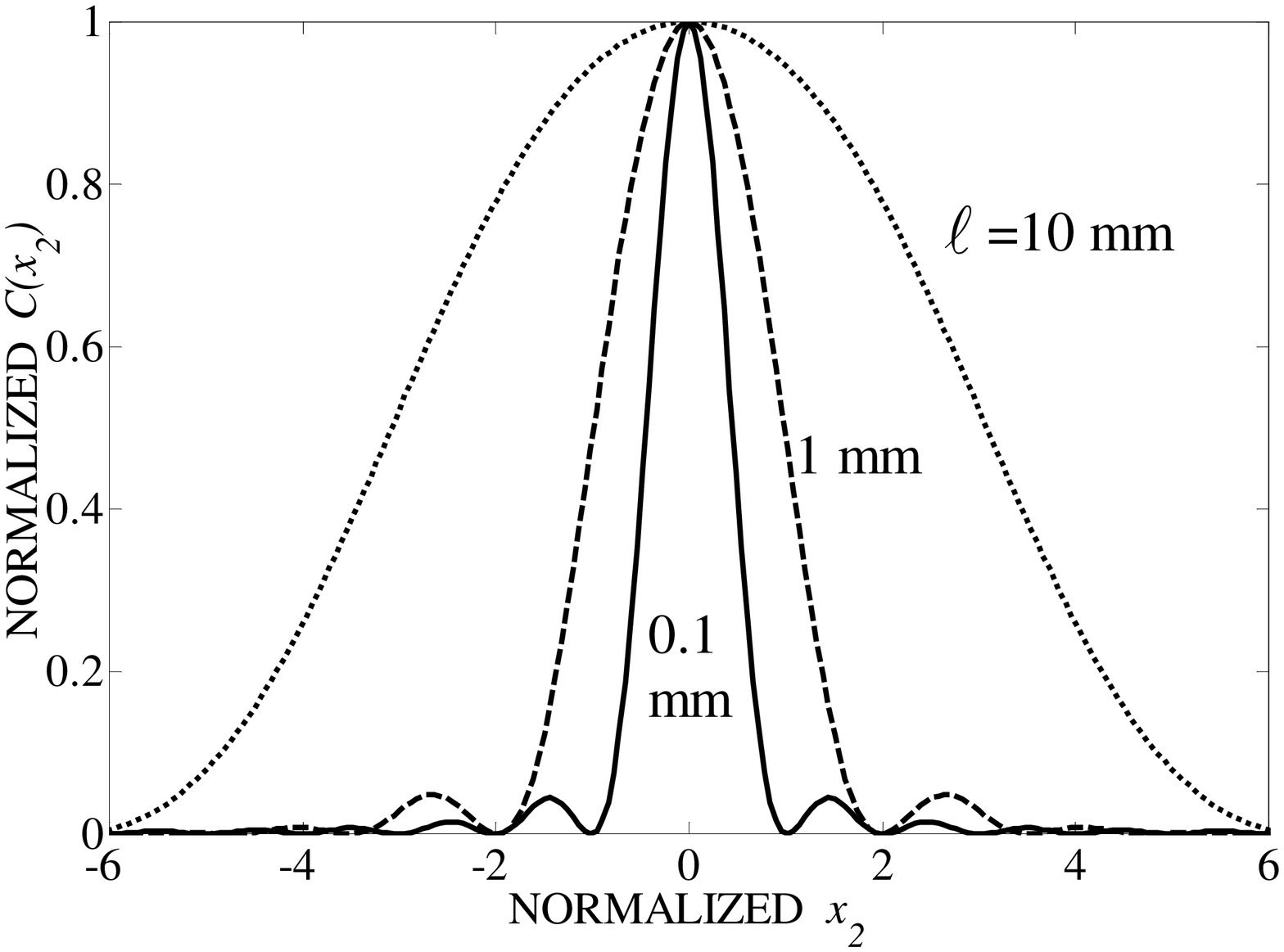}
 \end{figure}
\vskip6cm Figure 9, A. F. Abouraddy

\begin{figure}[h]
 \epsfxsize=6.8in \epsfysize=5.2in
 \epsffile{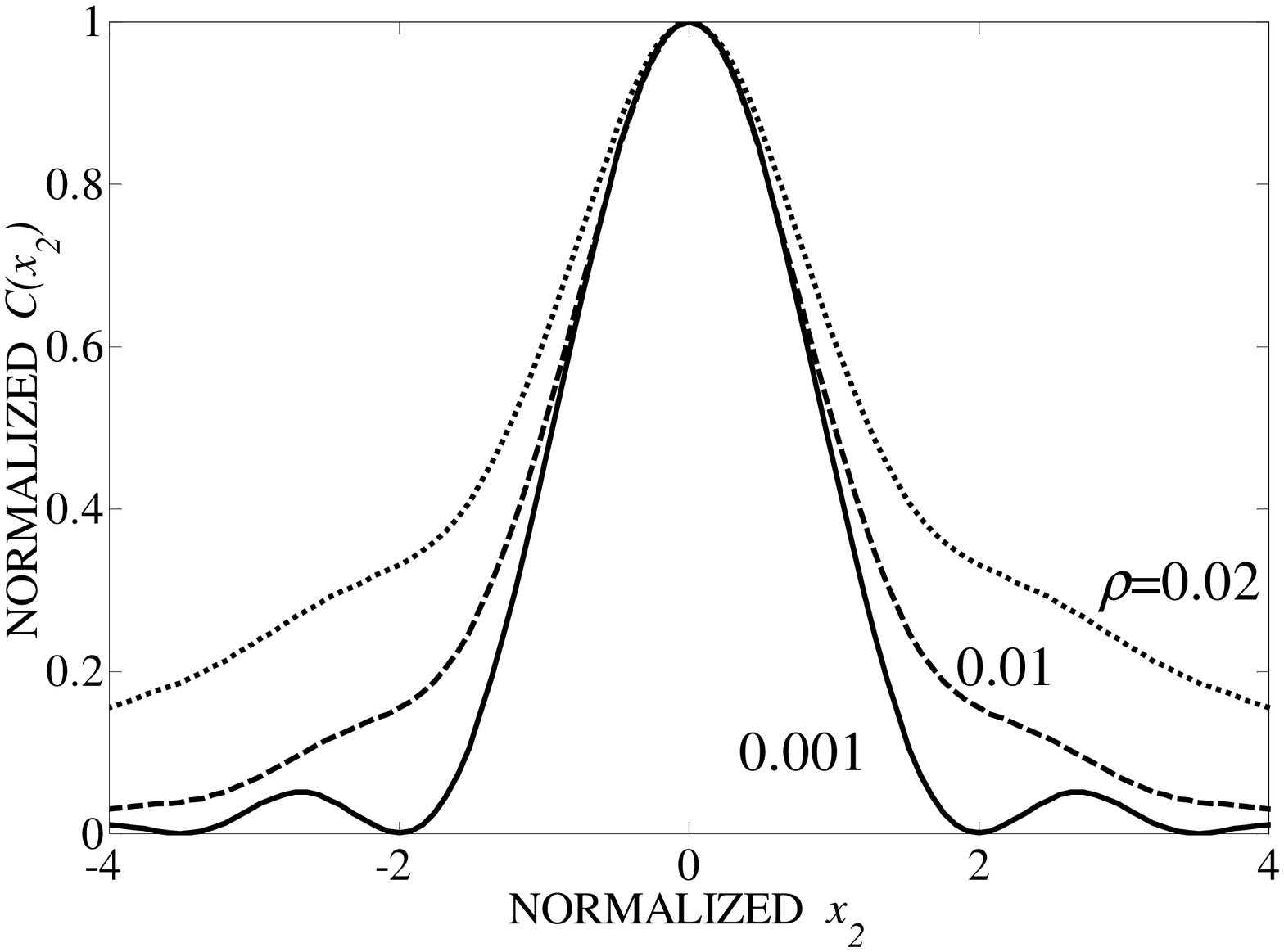}
 \end{figure}
\vskip6cm Figure 10, A. F. Abouraddy

\begin{figure}[h]
 \epsfxsize=6.8in \epsfysize=5.2in
 \epsffile{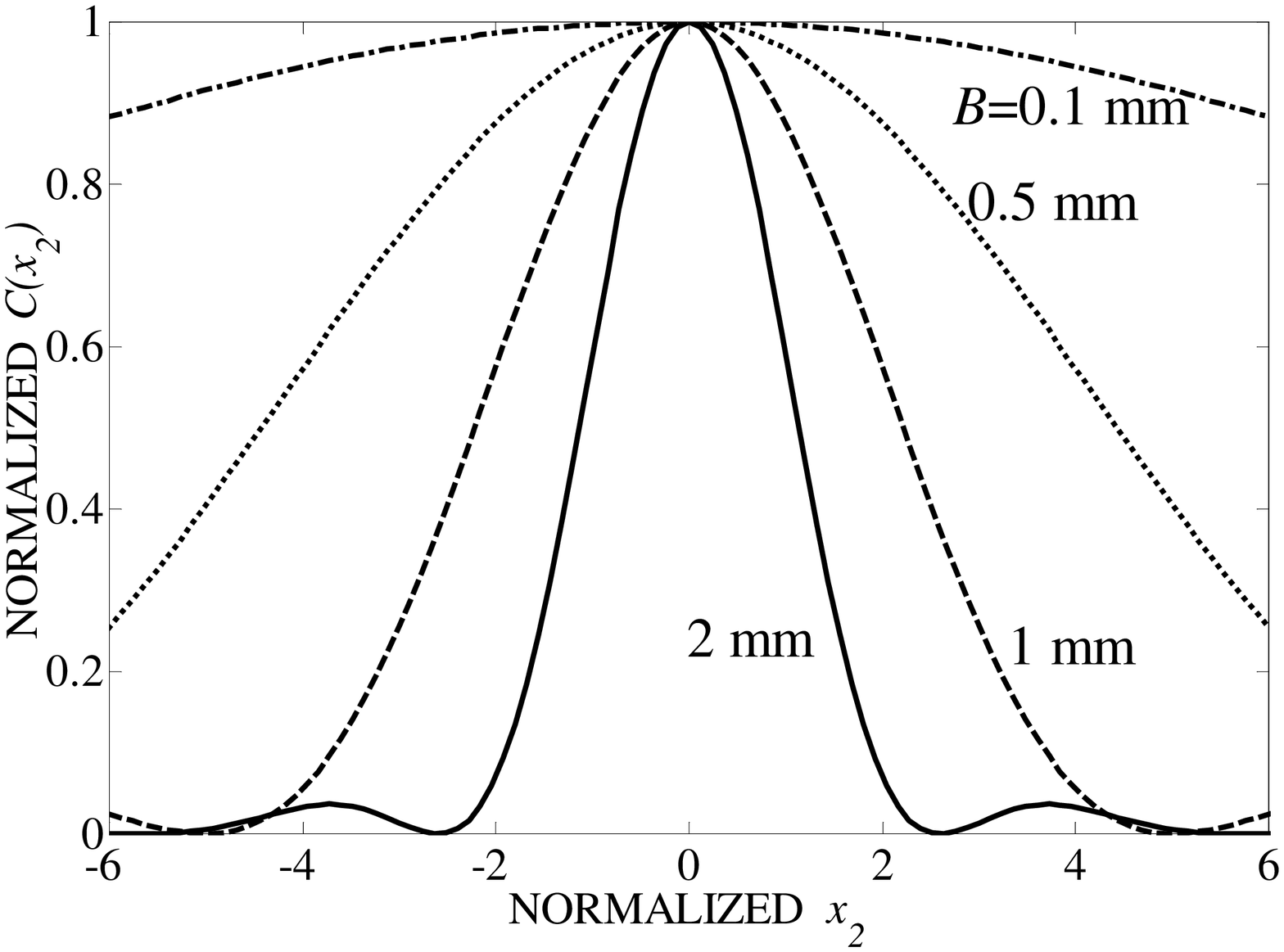}
 \end{figure}
\vskip6cm Figure 11, A. F. Abouraddy

\begin{figure}[h]
 \epsfxsize=6.8in \epsfysize=5.2in
 \epsffile{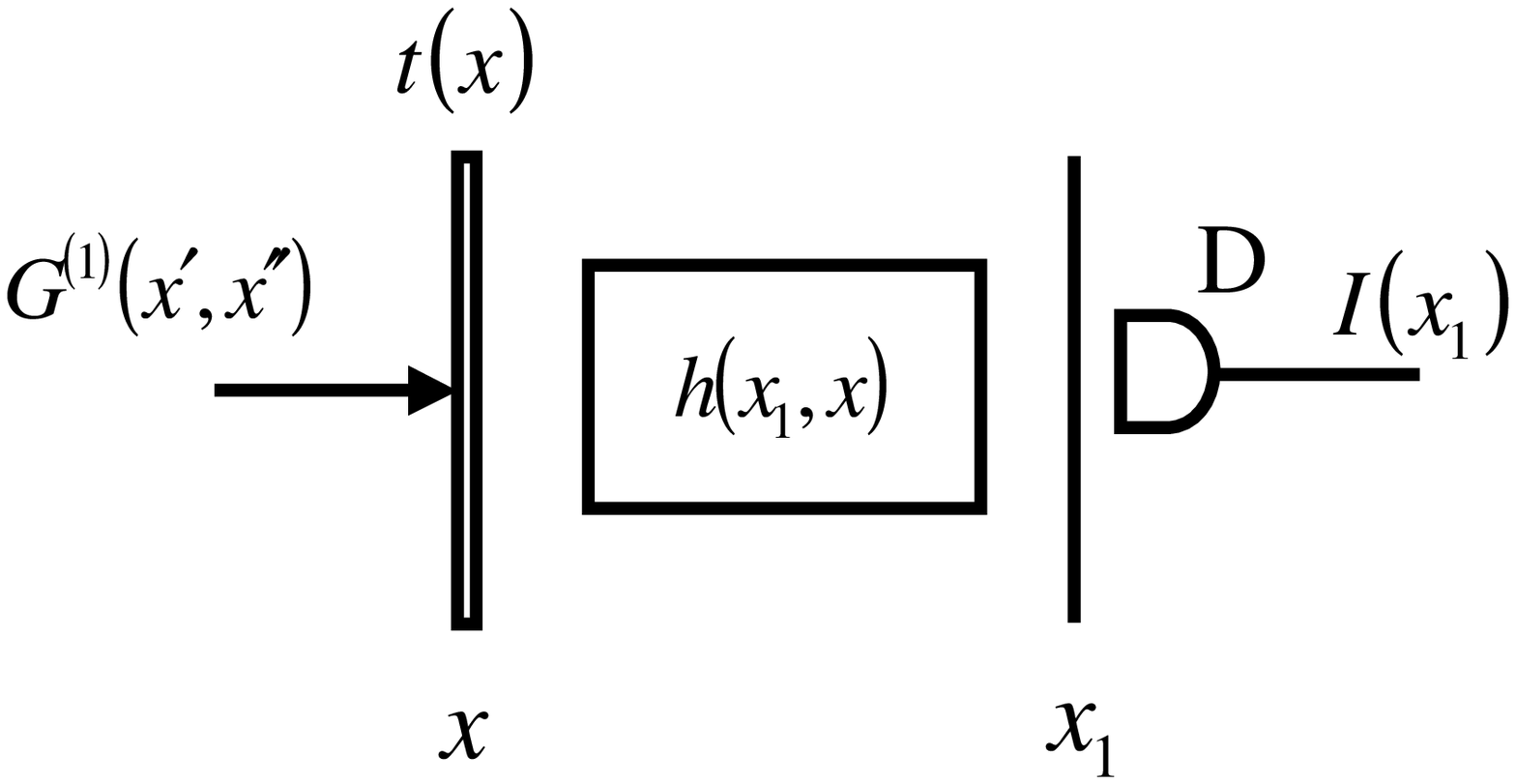}
 \end{figure}
\vskip6cm Figure 12, A. F. Abouraddy

\end{document}